\documentclass{JHEP}

\title{Integrability, Seiberg--Witten Models and Picard--Fuchs Equations}
\author{Jos\'e M. ISIDRO \\ Dipartimento di Fisica 
``G. Galilei'' -- Istituto Nazionale di Fisica Nucleare\\ Universit\`a 
degli Studi di  Padova,
Via Marzolo, 8 -- 35131 Padova, Italy,\\
\centerline{and}
Department of Physics, Brandeis University,  Waltham, MA 02454-9110, USA.
\email{isidro@pd.infn.it}}

\abstract{Expanded version of the author's contribution to the Concise 
Encyclop\ae dia of 
Supersymmetry, eds. Jonathan Bagger, Steven Duplij and Warren Siegel.}

\preprint{BRX-TH/483, DFPD00/TH/53, US-FT/22-00}


\begin{document}

\section{Introduction}\label{intro}

Seiberg--Witten (SW) theories \cite{SW} have, in their Coulomb branch, 
a moduli space of physical vacua which coincides with the moduli space 
of a certain class of Riemann surfaces $\Sigma$. In their simplest examples, 
the latter are hyperelliptic. Such is the situation when one considers a simple, 
classical gauge group, with or without matter hypermultiplets in the fundamental 
representation \cite{KLEMMREV}. An essential ingredient is the SW differential 
$\lambda_{SW}$ and its period integrals along (a subset of) the homology cycles 
of $\Sigma$. A knowledge of these period integrals amounts to the complete solution 
of the effective theory, as given by the full quantum prepotential ${\cal F}$ 
\cite{SW}. Different approaches have been undertaken in order to compute 
${\cal F}$ exactly, including instanton corrections \cite {MATONE, KLEMM}. 
One of them makes use of a set of partial differential equations, 
with respect to the moduli, satisfied by the periods of $\lambda_{SW}$ \cite{KLEMM}. 
Such differential equations are second--oder when the matter hypermultiplets 
are massless, while they become third--order when the masses are non--zero 
\cite{BRANDEIS}. In either case they can be traced back to a first--order system 
of differential equations satisfied by the periods of the holomorphic and meromorphic 
differentials of the second kind, called Picard--Fuchs (PF) equations \cite{GH}. 
It is therefore interesting to study the properties of such a first--order system 
of differential equations, without regard to their particular physical 
or mathematical origin \cite{ME}. 

For simplicity we will limit our discussion to the case of hyperelliptic 
Riemann surfaces. A systematic treatment of instanton corrections for 
hyperelliptic SW models has been given in \cite{DPINST}; see also 
\cite{BELLISAI, SANTIAGO}. Non--hyperelliptic Riemann surfaces arise naturally in 
the context of SW models when treated from an M--theoretic point of view
\cite{WITT}. The corresponding instanton corrections have been successfully studied 
in \cite{SCHNITZER}. Finally, non--hyperelliptic PF equations have also been analysed 
in \cite{ME}.

Throughout our study, $u_i$ will denote an arbitrary modulus. Thus, 
{\it e.g.}, in the context of SW models, specifying a value for the $u_i$ is 
equivalent to determining a physical vacuum state of a certain effective 
$N=2$ supersymmetric Yang--Mills theory. However, our analysis would hold just 
as well for the variation of, say, a complex structure on $\Sigma$.

\section{Hyperelliptic PF Equations}\label{hyper}

Let $p(x)$ be the complex polynomial
\begin{equation}
p(x)=\prod_{l=1}^{2g+1}(x-e_l)=\sum_{j=0}^{2g+1} s_j x^{2g+1-j},
\label{eq:1}
\end{equation}
where $g\geq 1$.  The
discriminant $\Delta$ of $p(x)$ is defined by 
\begin{equation}
\Delta=\prod_{l<n}^{2g+1}(e_l-e_n)^2.
\label{eq:2}
\end{equation}
Assume that $\Delta\neq 0$. Then the equation 
\begin{equation}
y^2=p(x)
\label{eq:3}
\end{equation}
defines a family of non--singular hyperelliptic Riemann surfaces $\Sigma$ 
of genus $g$. Each one of them is a twofold  covering of the Riemann sphere
branched  over the $e_l$, plus over the point at infinity.  
Choices of the $e_l$ such that $\Delta=0$ will produce singular surfaces. 
We will assume $\Delta(s_j)\neq 0$ in what follows.

The differential 1-forms on $\Sigma$
\begin{equation}
\omega_n=x^n\,{{\rm d} x\over y}, \qquad n=0,1,2, \ldots
\label{eq:4}
\end{equation}
are holomorphic for $0\leq n\leq g-1$, while they are meromorphic with 
vanishing
residues for all $n\geq g$. Let $\gamma\in  H_1(\Sigma)$ be an 
arbitrary 1-cycle.  The
period integrals
\begin{equation}
\Omega_n(\gamma)=\int_{\gamma}\omega_n,
\label{eq:5}
\end{equation}
and the differential equations they satisfy, will be our focus of 
attention. In order to derive the latter, a generalisation of equations (\ref{eq:4}) 
and (\ref{eq:5}) is needed. Let ${\bf Z}^+$ denote the non-negative integers 
and ${1\over 2} {\bf Z}^-$ the negative half-integers.
Consider $\mu\in {1\over 2} {\bf Z}^-$  and $n\in {\bf Z}^+$, and let a 
given 1-cycle $\gamma\in
H_1(\Sigma)$ be fixed. Define the $\mu$-period of $x^n$ along 
$\gamma$, denoted by
$\Omega_n^{(\mu)}(\gamma)$,  as
\begin{equation}
\Omega_n^{(\mu)}(\gamma)=(-1)^{\mu +1}\Gamma  (\mu + 1) 
\int_{\gamma}{x^n\over p^{\mu + 1}(x)}\,{\rm d} x,
\label{eq:6}
\end{equation}
where $\Gamma$ stands for Euler's gamma function. One can prove 
that the $\Omega_n^{(\mu)}(\gamma)$ are well defined as a function of the
homology class of $\gamma$, and that they satisfy the
following recursion relations \cite{ME}:
\begin{equation}
\Omega_n^{(\mu)}(\gamma)={1\over n+1-(1+\mu)(2g+1)}\sum_{j=0}^{2g+1} j s_j
\Omega_{n+2g+1-j}^{(\mu +1)}(\gamma)
\label{eq:7}
\end{equation}
and
\begin{equation}
\Omega_n^{(\mu+1)}(\gamma)={1\over n-2g-(1+\mu)(2g+1)}\sum_{j=1}^{2g+1} 
\big[(1+\mu)(2g+1-j)-(n-2g)\big] s_j \Omega_{n-j}^{(\mu +1)}(\gamma).
\label{eq:8}
\end{equation}

We define the {\it basic range} $R$ to be the set of all 
$n\in {\bf Z}^+$ such that $0\leq n\leq
2g-1$. Let $Z[s_j]$ be the ring of polynomials in the $s_j$ with 
complex coefficients. The set of all periods $\Omega_n^{(\mu+1)}(\gamma)$ 
as $n$ runs over ${\bf Z}^+$, when both $\mu$ and $\gamma$ are kept fixed, 
defines a module over $Z[s_j]$; let it be denoted by ${\cal
M}^{(\mu+1)}(\gamma)$.
Then equations (\ref{eq:7}) and (\ref{eq:8}) prove that ${\hbox {dim}
}\,{\cal M}^{(\mu+1)}(\gamma)=2g$. This  follows from the observation
that equation (\ref{eq:8}) will express any
$\Omega_n^{(\mu+1)}(\gamma)$, where $n\geq 2g$, as a certain linear 
combination of
$\Omega_m^{(\mu+1)}(\gamma)$, with $m<n$, and with coefficients that 
will be certain
homogeneous polynomials in the $s_j$. Repeated application of this 
recursion will eventually allow
to reduce all terms in that linear combination to a sum over the 
periods of the basic range $R$.
The $\Omega_n^{(\mu+1)}(\gamma)$, where $n\in R$, will be called 
{\it basic periods} of ${\cal M}^{(\mu+1)}(\gamma)$.

The two modules ${\cal M}^{(\mu)}(\gamma)$ and ${\cal M}^{(\mu+1)}
(\gamma)$ are clearly
isomorphic. We can make this isomorphism more explicit as follows. 
Let $\Omega_n^{(\mu)}(\gamma)$
and $\Omega_m^{(\mu+1)}(\gamma)$, where both $n$ and $m$ run over $R$, 
be  bases
of ${\cal M}^{(\mu)}(\gamma)$ and ${\cal M}^{(\mu+1)}(\gamma)$, 
respectively.
Arrange them as column vectors $\Omega^{(\mu)}=(\Omega_0^{(\mu)},
\Omega_1^{(\mu)},\ldots, \Omega_{2g-1}^{(\mu)})^t$ and 
$\Omega^{(\mu+1)}=(\Omega_0^{(\mu+1)},
\Omega_1^{(\mu+1)},\ldots, \Omega_{2g-1}^{(\mu+1)})^t$. For notational 
simplicity we have suppressed
the dependence on $\gamma$. Now, for every $\mu\in {1\over 2} {\bf Z}^-$ 
there exists a unique  matrix
$M^{(\mu)}$ such that
\begin{equation}
\Omega^{(\mu)}=M^{(\mu)}\Omega^{(\mu+1)}.
\label{eq:15}
\end{equation}
The entries of $M^{(\mu)}$ are certain homogeneous polynomials in the 
$s_j$. $M^{(\mu)}$ is
non-singular when $\Delta(s_j)\neq 0$. A proof of the statement (\ref{eq:15}) 
can be found in \cite{ME}. 

As a function of the $s_j$, we have  that ${\rm det} \,M^{(\mu)}$ 
can only vanish when
$\Delta(s_j)=0$. The entries of $M^{(\mu)}$ are polynomials in the 
$s_j$, hence
${\rm det} \,M^{(\mu)}$ will also be a polynomial in the $s_j$. 
Decompose $\Delta(s_j)$ into
irreducible factors.  Up to an overall complex constant, 
${\rm det} \,M^{(\mu)}$ must therefore
decompose as a product of exactly those same factors present in the 
decomposition of
$\Delta(s_j)$, possibly with different multiplicities (eventually with 
zero multiplicity, {\it i.e.},
$\Delta(s_j)$ might have more zeroes than ${\rm det} \,M^{(\mu)}$).

Next we consider derivatives of periods. We have 
\begin{equation}
{\partial \Omega_n^{(\mu)}\over\partial s_j}=\Omega_{n+2g+1-j}^{(\mu+1)}.
\label{eq:16}
\end{equation}
Now take $n\in R$ in equation (\ref{eq:16}), and use the recursion
relation (\ref{eq:8})  as many times as necessary in
order to pull the subindex $n+2g+1-j$ back into $R$. As $n$ runs over 
$R$, the right-hand side of
equation (\ref{eq:16}) defines the rows of a $(2g\times 2g)$-dimensional
matrix, 
$D^{(\mu)}_j$. In matrix
form, equation (\ref{eq:16}) reads
\begin{equation}
{\partial \Omega^{(\mu)}\over\partial s_j}=D^{(\mu)}_j \,\Omega^{(\mu+1)}.
\label{eq:17}
\end{equation}
Since equation (\ref{eq:15}) can be inverted when $\Delta(s_j)\neq 0$,
we have
\begin{equation} 
{\partial \Omega^{(\mu)}\over\partial s_j}=D^{(\mu)}_j \,(M^{(\mu)})^{-1}\,
\Omega^{(\mu)} = S^{(\mu)}_j\,
\Omega^{(\mu)},
\label{eq:18}
\end{equation}
where we have defined $S^{(\mu)}_j=D^{(\mu)}_j \,(M^{(\mu)})^{-1}$. 
We finally set $\mu=-1/2$
in order to obtain the PF equations of the hyperelliptic Riemann 
surface $\Sigma$:
\begin{equation}  
{\partial \over\partial s_j}\Omega^{(-1/2)}= S^{(-1/2)}_j\,\Omega^{(-1/2)}.
\label{eq:19}
\end{equation}
They express the derivatives of the basic periods with respect to the 
$s_j$, as certain linear
combinations of the same basic periods. The $4g^2$ entries of the 
matrix $S^{(-1/2)}_j$
are certain rational functions of the $s_j$, explicitly computable 
using the recursion relations  
above. Finally, from a knowledge of the coefficients $s_j$ as functions 
of the moduli $u_i$,
application of the chain rule and equation (\ref{eq:19}) produces the
desired  PF equations $\partial\Omega/\partial u_i$.

\section{The Inverse PF Equations and Integrability}\label{irs}

\subsection{The inverse PF equations}\label{irsi}

Dropping a total derivative under the integral along $\gamma\in H_1(\Sigma)$
we have, from equation (\ref{eq:3}), that
\begin{equation}
n{x^{n-1}\over y} = {x^n\over 2 y^3}\, p'(x), \qquad n\geq 0.
\label{eq:20}
\end{equation}
Next we define, for any non--negative integer $k$, the differential $\lambda_k$
\begin{equation}
\lambda_k={x^k\over y}\, p'(x)\, {\rm d} x, \qquad k\geq 0.
\label{eq:21}
\end{equation}
Modulo total derivatives it holds that
\begin{equation}
{\partial \lambda_k\over \partial s_l}= (-1)^{l+1}\, k\, 
{x^{k+2g-l}\over y}\, {\rm d} x.
\label{eq:22}
\end{equation}
It is tempting to call equation (\ref{eq:22}) the {\it potential 
property}. In fact it is strongly reminiscent of the property of the
SW differential $\lambda_{SW}$, that its modular derivatives 
are holomorphic diferentials on $\Sigma$ \cite{KLEMMREV}:
\begin{equation}
{\partial\lambda_{SW}\over\partial u_i}= \sum_{j=0}^{g-1}c_i^j\, 
\omega_j,
\label{potential}
\end{equation}
where the $\omega_j$ are as in equation (\ref{eq:4}), and $c_i^j$ is a 
constant matrix.
We will return to this point in section \ref{conn}. 
One can now easily prove that the period integrals of $\lambda_k$ along any 
$\gamma\in H_1(\Sigma)$ are well--defined functions of the homology class 
of $\gamma$, and that taking modular derivatives does not alter this 
property. For the second derivatives of $\lambda_k$ we find
\begin{equation}
{\partial^2\lambda_k\over \partial s_j\partial s_l}= (-1) ^{j+l}\,k\, 
{x^{k+4g +1- j-l}\over 2y^3}.
\label{eq:23}
\end{equation}
Then it is easy to prove that the operators
\begin{equation}
{\cal L}_l^{(k)}=(2g+1+k-l){\partial\over \partial s_l}+
\sum_{j=0}^{2g+1}(2g+1-j)\, s_j\, {\partial^2\over \partial s_j\partial s_l}
\label{eq:24}
\end{equation}
annihilate the periods of $\lambda_k$. The proof follows from 
equations (\ref{eq:20}), (\ref{eq:22}) and (\ref{eq:23}).
With the notation of equation (\ref{eq:4}), the potential property (\ref{eq:22}) 
reads
\begin{equation}
{\partial\lambda_k\over \partial s_l}=(-1)^{l+1}\,k\, \omega_{2g+k-l},
\label{eq:26}
\end{equation}
and for any fixed value of  $k\geq 1$, the basic range $R$ of differentials is 
spanned by those values of $l$ such that $1\leq l-k\leq 2g$. 
If we now denote by $\Omega_n$ the period integrals of the above 
$\omega_n$, then these $\Omega_n$ correspond to the $\Omega_n^{(-1/2)}$ of 
section \ref{hyper}. Finally it is straightforward to establish, using equations 
(\ref{eq:24}) and (\ref{eq:26}), that the following equations hold:
\begin{equation}
\Omega_{2g+k-l}={-1\over 2g+1+k-l} 
\sum_{j=0}^{2g+1}(2g+1-j)\,s_j\,{\partial\over\partial s_j}\Omega_{2g+k-l},
\label{eq:27}
\end{equation}
or, equivalently,
\begin{equation}
\Omega_{2g+k-l}={(-1)^{l+1}\over 
2g+1+k-l}\sum_{j=0}^{2g+1}(-1)^j\,(2g+1-j)\,s_j\,{\partial\over\partial 
s_l}\Omega_{2g+k-j}.
\label{eq:28}
\end{equation}

\subsection{Connection with integrability}\label{conn}

Let us summarise our results so far. In section \ref{hyper} 
we derived a set of equations for the derivatives of periods 
as linear combinations (with moduli--dependent coefficients) 
of periods. Conversely, in section \ref{irsi} we have expressed 
periods as linear combinations of derivatives of periods. 

The purpose of this section, however, goes beyond the triviality of 
inverting the PF equations (\ref{eq:19}). Rather, we would like to 
relate the potential property (\ref{potential}) of the SW differential 
with the notion of integrability in the context of SW models 
\cite{DONAGIWITTEN, MOROZOV, CARROLL}. The integrability of SW models 
can be traced back to the existence of a prepotential ${\cal F}$.
In principle, the latter can be obtained exactly by integrating the PF equations. 
We recall, however, that the PF equations for SW models are at least 
second--order and that, in their derivation, one makes crucial use of 
the potential property (\ref{potential}) of the SW differential $\lambda_{SW}$ 
\cite{BRANDEIS}. In actual SW models, modular derivatives 
of $\lambda_{SW}$ need not generate the full $g$-dimensional space of 
holomorphic differentials on $\Sigma$, the reason being that the Weyl group
may select a particular subspace. Once this Weyl symmetry is taken 
into account, the rank of the matrix $c_i^j$ in equation (\ref{potential}) 
equals the rank of the gauge group.

On the contrary, the PF equations dealt with in sections \ref{hyper} and 
\ref{irsi} here are first--order. They apply to {\it any}\/ hyperelliptic 
Riemann surface $\Sigma$. When written as in section \ref{hyper}, 
no use is made of the property (\ref{eq:22}). In general, an arbitrary 
hyperelliptic Riemann surface such as (\ref{eq:3}) does {\it not} possess 
a SW differential, in the sense of equation (\ref{potential}).

We can however state two necessary and sufficient conditions for the Riemann 
surface $\Sigma$ to be integrable, in the sense of SW models. First, 
an $n$-dimensional subspace of holomorphic differentials $\omega_j$ must be 
derivable from a {\it potential} $\lambda$. That is, there must exist 
a (meromorphic) differential $\lambda$, and a subset of $m\leq 2g-1$ independent 
moduli $u_l$ \footnote{The moduli space of hyperelliptic Riemann surfaces is 
$(2g-1)$-dimensional \cite{SCHLICHENMAIER}.}, such that the equation
\begin{equation}
{\partial\lambda\over\partial u_l}=\sum_{j=0}^{g-1} d_l^j \omega_j
\label{cosncoeff}
\end{equation}
holds, for a certain rank-$n$ matrix of constant coefficients $d_l^j$.
Second, when susbtituting equation (\ref{cosncoeff}) into the first--order 
PF equations (\ref{eq:19}), the latter must be completely expressible 
in terms of the $u_l$. This will lead to a system of PF equations which 
will be second--order {\it and}\/ integrable, {\it i.e.}, it will define 
a prepotential ${\cal F}$. 
 
\section{Further Applications}\label{further}

The reader may want to consult \cite{AG, KETOV} for nice introductions to 
SW models. Especially \cite{BILAL} takes the approach based on PF equations.
Apart from SW theories, other recent physical contexts in which hyperelliptic period 
integrals and PF equations play a role include special geometry in $N=2$ supersymmetry 
\cite{CERESOLE}, mirror symmetry \cite{THEISEN, VOISIN}, Landau--Ginzburg 
models, topological field theory and Calabi--Yau manifolds \cite{FRE}, 
integrability in connection with $N=2$ supersymmetric Yang--Mills theory 
\cite{DHOKERPHONG, LNS} and Donaldson--Witten theory \cite{MAS}, 
and the WDVV equation \cite{MIRONOV, MEWDVV}.
In the mathematical literature, PF equations govern the variation of a 
Hodge structure on a manifold \cite{GH}. They also arise in singularity theory 
\cite{ARNOLD, YAKOVENKO} and noncommutative geometry \cite{CONNES}, 
in connection with Hilbert's 16th and 21st problems, respectively.

\acknowledgments

It is a great pleasure to thank Diego Bellisai, Gaetano Bertoldi, Jos\'e D.
Edelstein, Kurt Lechner, Pieralberto Marchetti, Marco Matone, Jo\~ao P. Nunes, 
Paolo Pasti, Howard J. Schnitzer, Dmitri Sorokin, Mario Tonin, Niclas Wyllard 
and J.-G. Zhou for interesting discussions. The author would like to thank the Department 
of Physics of Brandeis University, where the final part of this work was completed, 
for hospitality. This work has been supported by a Fellowship from Istituto Nazionale 
di Fisica Nucleare (Italy).

\end{document}